\documentclass[aps,prd,twocolumn,showpacs,preprintnumbers,nofootinbib,amsmath,amssymb]{revtex4}

\usepackage{graphicx}
\usepackage{epstopdf}
\usepackage{dcolumn}
\usepackage{bm}
\usepackage{amsmath}
\usepackage[normalem]{ulem}
\usepackage{blindtext}
\usepackage{subfigure}
\usepackage{color}

\newcommand{\beq}{\begin{eqnarray}}
\newcommand{\eeq}{\end{eqnarray}}

\def\spose#1{\hbox to 0pt{#1\hss}}
\def\ltapprox{\mathrel{\spose{\lower 3pt\hbox{$\mathchar"218$}}
\raise 2.0pt\hbox{$\mathchar"13C$}}}

\begin{document}

\title{Magnetic field effects on the static quark potential at zero 
and finite temperature}
\author{Claudio Bonati}
\email{claudio.bonati@df.unipi.it}
\affiliation{
Dipartimento di Fisica dell'Universit\`a
di Pisa and INFN - Sezione di Pisa,\\ Largo Pontecorvo 3, I-56127 Pisa, Italy}

\author{Massimo D'Elia}
\email{massimo.delia@unipi.it}
\affiliation{
Dipartimento di Fisica dell'Universit\`a
di Pisa and INFN - Sezione di Pisa,\\ Largo Pontecorvo 3, I-56127 Pisa, Italy}

\author{Marco Mariti}
\email{mariti@df.unipi.it}
\affiliation{
Dipartimento di Fisica dell'Universit\`a
di Pisa and INFN - Sezione di Pisa,\\ Largo Pontecorvo 3, I-56127 Pisa, Italy}

\author{Michele Mesiti}
\email{michele.mesiti@pi.infn.it}
\affiliation{
Dipartimento di Fisica dell'Universit\`a
di Pisa and INFN - Sezione di Pisa,\\ Largo Pontecorvo 3, I-56127 Pisa, Italy}

\author{Francesco Negro}
\email{fnegro@pi.infn.it}
\affiliation{
Dipartimento di Fisica dell'Universit\`a
di Pisa and INFN - Sezione di Pisa,\\ Largo Pontecorvo 3, I-56127 Pisa, Italy}

\author{Andrea Rucci}
\email{andrea.rucci@pi.infn.it}
\affiliation{
Dipartimento di Fisica dell'Universit\`a
di Pisa and INFN - Sezione di Pisa,\\ Largo Pontecorvo 3, I-56127 Pisa, Italy}

\author{Francesco Sanfilippo}
\email{f.sanfilippo@soton.ac.uk}
\affiliation{School of Physics and Astronomy, University of Southampton, SO17 1BJ Southampton, 
United Kindgdom}

\date{\today}

\begin{abstract}
We investigate the static $Q\bar{Q}$ potential at zero and finite
temperature in the presence of a constant and uniform external
magnetic field $\vec{B}$, for several values of the lattice spacing
and for different orientations with respect to $\vec{B}$. As a
byproduct, we provide continuum limit extrapolated results for the
string tension, the Coulomb coupling and the Sommer parameter at $T =
0$ and $B = 0$.
We confirm the presence in the continuum of a $B$-induced anisotropy,
regarding essentially the string tension, for which it
is of the order of 15\% at $|e| B \sim 1~{\rm GeV}^2$ and would suggest, if extrapolated to larger fields, a
vanishing string tension along the magnetic field for $|e| B \gtrsim
4$ GeV$^2$.
The angular dependence for $|e| B \lesssim 1$ GeV$^2$ can be nicely
parametrized by the first allowed term in an angular Fourier
expansion, corresponding to a quadrupole deformation.  Finally, for $T
\neq 0$, the main effect of the magnetic field is a general
suppression of the string tension, leading to a precocious loss of the
confining properties: this happens even before the appearance of
inverse magnetic catalysis in the chiral condensate, supporting the idea 
that the
influence of the magnetic field on the confining properties is 
the leading effect originating the decrease of $T_c$ as a function of $B$.
\end{abstract}


\maketitle

\section{Introduction}
\label{intro}

In the Standard Model, the strong and the electroweak sectors are
connected by quarks, which are subject to both type of
interactions. In general, electroweak interactions are expected to
induce relatively small corrections to strong interaction dynamics,
however this may be not true in the presence of very strong
electromagnetic backgrounds, with field values comparable to the QCD
scale. This is a situation which is relevant to many contexts, ranging
from non-central heavy ion collisions~\cite{hi1, hi2, hi3, hi4,
  tuchin, Holliday:2016lbx} and the cosmological electroweak phase
transition~\cite{vacha,grarub}, with magnetic fields reaching or
exceeding $10^{16}$ Tesla ($e B \sim 1$~GeV$^2$), to
magnetars~\cite{magnetars}, where magnetic fields are expected to be
of the order of $10^{11}$ Tesla on the surface but could be
significantly larger in the inner cores.

How strong interactions get modified by such large magnetic fields has
been the subject of many recent theoretical studies (see, e.g.,
Refs.~\cite{Kharzeev:2012ph, Miransky:2015ava} for reviews), with lattice
simulations representing a viable and effective tool to explore the
issue starting from the first principles of QCD. One important feature
is that gluon fields, even if not directly coupled to electromagnetic
fields, undergo significant modifications, through effective QED-QCD
interactions induced by quark loops, as can be seen both by lattice
simulations \cite{D'Elia:2010nq, Buividovich:2010tn,D'Elia:2011zu, Bali:2011qj,
  Ilgenfritz:2012fw,Bali:2012zg, D'Elia:2012zw, Bruckmann:2013oba,Ilgenfritz:2013ara,Bonati:2014ksa,Schafer:2015wja,D'Elia:2015dxa,Bonati:2013vba,Levkova:2013qda,Bali:2013esa,Bali:2014kia,Cea:2015ifa}
and by analytical studies using several approaches, which range from
perturbation theory to effective field theories, from NJL models to
functional renormalization group techniques \cite{Musakhanov:1996cv,
Rafelski:1998tc, Elze:1998wm, Chernodub:2010bi, Asakawa:2010bu,
Galilo:2011nh, Andreichikov:2012xe, Kojo:2012js, Kojo:2013uua,
Watson:2013ghq, Andersen:2013swa, Ozaki:2013sfa, Kamikado:2013pya,
Mueller:2014tea, Mueller:2015fka,
Fraga:2012fs,Fraga:2012ev,Rougemont:2015oea,Dudal:2015wfn,Farias:2016gmy,
Finazzo:2016mhm,McInnes:2015kec}. 
A striking consequence of the
QED-QCD coupling is the distortion of the zero temperature static
quark-antiquark potential. 
Results reported in
Ref.~\cite{Bonati:2014ksa} showed the emergence of anisotropies both
in the linear part (string tension) and in the Coulomb part of the
potential. This behaviour is consistent with the results of some of
the existing model computations \cite{Miransky:2002rp,
  Chernodub:2014uua, Rougemont:2014efa, Ferrer:2014qka,
  Simonov:2015yka} and may have relevant phenomenological
consequences, especially for the spectrum of heavy quark bound
states~\cite{Alford:2013jva,Dudal:2014jfa, Cho:2014loa, Taya:2014nha, Bonati:2015dka,
  Hattori:2015aki, Suzuki:2016kcs,Guo:2015nsa,Fukushima:2015wck,Gubler:2015qok}.

The purpose of this study is to move one step forward in our
comprehension of magnetic-induced effects on non-perturbative QCD
dynamics. First of all, we present a refinement of our zero
temperature data following three different directions:
\begin{itemize}
\item a complete analysis of the angular dependence of the potential
  (in Ref.~\cite{Bonati:2014ksa} only quark-antiquark separations
  parallel and orthogonal to the magnetic field were analyzed). This
  is important to allow for a realistic modelling of the
  quark-antiquark potentials to be used in the computation of the
  spectrum of heavy quarkonia;

\item the inclusion of new simulations on finer lattices will permit
  us to extrapolate results reported in Ref.~\cite{Bonati:2014ksa} to
  the continuum limit;

\item the investigation in the regime of fields significantly larger
  than those used in Ref.~\cite{Bonati:2014ksa}, in order to inquire
  whether new unexpected phenomena may take place in the QCD vacuum
  under the influence of extremely strong background fields.
\end{itemize}

As a byproduct of this investigation, we will provide a continuum
extrapolation of the static quark-antiquark potential with physical
quark masses also for the standard case of vanishing magnetic field.

In the second part of our study, we investigate how the effect of the
magnetic field on the static potential gets modified by the
temperature, a step which is important for a full comprehension of the
properties of the thermal medium in the presence of magnetic
backgrounds. In this case the static potential (more correctly the
free energy) is extracted from Polyakov loop correlators in place of
Wilson loop expectation values used at $T=0$. For temperatures below
the pseudo-critical temperature $T_c\sim 155\,\mathrm{MeV}$ (at which
chiral symmetry gets restored and quark and gluon degrees of freedom
deconfine), one still expects that heavy quark-antiquark interactions
can be described in terms of a confining potential, with a string
tension which goes to zero as $T_c$ is approached. Two main
questions will be addressed by our study in this regime: {\em i)} does
the anisotropy survive also in the finite temperature case?  {\em ii)}
does the magnetic field enhance the suppression of the string tension,
meaning that a phenomenon similar to inverse magnetic 
catalysis~\cite{Bali:2012zg}, observed for chiral symmetry, 
takes place also for
the confining properties of the medium?

The paper is organized as follows.  In Section~\ref{numsetup} we illustrate the
setup of our numerical simulations and the techniques adopted to extract the
static potential, both at zero and at finite temperature; in 
Section~\ref{results}
we present our numerical results and finally, in Section~\ref{concl}, we draw
our conclusions.

\section{Numerical methods}
\label{numsetup}

The discretization of the $N_f=2+1$ QCD action adopted in this work is a
combination of the Symanzik tree-level improved gauge action and of stout
improved rooted staggered fermions. Explicitly, the partition function is
written as 
\begin{equation}\label{partfunc}
Z(B) = \int \!\mathcal{D}U \,e^{-\mathcal{S}_{Y\!M}}
\!\!\!\!\prod_{f=u,\,d,\,s} \!\!\!
\det{({D^{f}_{\textnormal{st}}[B]})^{1/4}}\ .
\end{equation}
Here $\mathcal{D}U$ is the functional integration over the $SU(3)$
link variables and $\mathcal{S}_{Y\!M}$ stands for the tree-level
improved action~\cite{weisz, curci}:
\begin{equation}\label{tlsyact}
\mathcal{S}_{Y\!M}= - \frac{\beta}{3}\sum_{i, \mu \neq \nu} \left(
\frac{5}{6} W^{1\!\times \! 1}_{i;\,\mu\nu} - \frac{1}{12}
W^{1\!\times \! 2}_{i;\,\mu\nu} \right),
\end{equation}
where the real part of the trace of the $1\!\times \! 1$ and
$1\!\times \!2$ loops is denoted by $W^{1\!\times \! 1}_{i;\,\mu\nu}$
and $W^{1\!\times \!  2}_{i;\,\mu\nu}$ respectively. Finally, the
staggered Dirac matrix
\begin{equation}\label{rmmatrix}
\begin{aligned}
(D^f_{\textnormal{st}})_{i,\,j} =\ & am_f
  \delta_{i,\,j}+\!\!\sum_{\nu=1}^{4}\frac{\eta_{i;\,\nu}}{2}
  \left(u^f_{i;\,\nu}U^{(2)}_{i;\,\nu}\delta_{i,j-\hat{\nu}}
  \right. \nonumber\\ &-\left. u^{f*}_{i-\hat\nu;\,\nu}U^{(2)\dagger}_{i-\hat\nu;\,\nu}\delta_{i,j+\hat\nu}
  \right)
\end{aligned}
\end{equation}
is written by using the two times stout-smeared links
$U^{(2)}_{i;\,\mu}$ \cite{morning} (with isotropic smearing parameter
$\rho=0.15$) and the $U(1)$ parallel transporters $u^f_{i;\,\mu}$,
where the $i$ index denotes the position in the lattice and the $\mu$
index denotes the direction of the link.

\begin{table}[t]
\centering
\begin{tabular}{ |c|c|c|c|c| }
\hline
lattice size &   $a [\textrm{fm}]$ & $\beta$   & $am_s$ & $b$ \\ \hline
$24^4$            & 0.2173(4)      & 3.55      & 0.1020 & 0,12,16,24,32,40   \\   
$32^4$            & 0.1535(3)      & 3.67      & 0.0639 & 0,12,16,24,32,40   \\   
$40^4$            & 0.1249(3)      & 3.75      & 0.0503 & 0,8,12,16,24,32,40 \\   
$48^3 \times 96$  & 0.0989(2)      & 3.85      & 0.0394 & 0,8,16,24,32   \\   
\hline
\end{tabular}
\caption{Simulation parameters for the $T=0$ runs, chosen according to
  Refs.~\cite{physline1, physline2} and corresponding to physical
values of the pion mass and of the 
strange-to-light mass ratio, 
$m_{s}/m_{u,d}=28.15$. The systematic error on $a$ is
  about $2 - 3~\%$~\cite{physline2, physline3}. }
\label{tab:paramT0}
\end{table}
 
For a magnetic field directed along the $\hat{z}$ direction, a
possible choice of the abelian transporters is ($q_f$ is the fermion
charge)
\begin{eqnarray}\label{bfield}
u^f_{i;\,y}=e^{i a^2 q_f B_{z} i_x} \ , \quad
{u^f_{i;\,x}|}_{i_x=L_x}=e^{-ia^2 q_f L_x B_z i_y}\, ,
\end{eqnarray}
with all the other $U(1)$ link variables being equal to 1.  Moreover,
for these transporters to describe a uniform magnetic field on the
lattice torus, the value of $B_z$ has to satisfy the following
quantization condition \cite{thooft, bound3, wiese, review}
\begin{equation}\label{bquant}
eB_z={6 \pi b_z}/{(a^2 N_x N_y)} \, ; \ \ \ \ \  b_z \in \mathbb{Z} \, .
\end{equation}
In the following we will consider also the case of a magnetic field
$\vec{B}$ not directed along one of the coordinate axes. In this case,
each component of $\vec{B}$ generates transporters analogous to
Eq.~\eqref{bfield} in the corresponding orthogonal plane, and the
final $U(1)$ phases appearing in the fermion matrix are the product of
the phases that would be generated by each component separately. All
the components have to satisfy a quantization condition analogous to
Eq.~\eqref{bquant}: as a consequence, the magnetic field on the
lattice can be represented by the vector $\vec{b}$ having integer
components. If $N_x$=$N_y$=$N_z$, i.e. if the quantization condition
in Eq.~\eqref{bquant} is the same for all components, then the
magnetic field $\vec B$ is parallel to $\vec b$.

The values of the bare parameters used in our simulations have been
chosen so as to move on a line of constant physics, corresponding to
physical values of the quark masses: to do that, we have followed the
determination reported in Refs.~\cite{physline1, physline2}. In
Tab.~\ref{tab:paramT0} we list for convenience the values of the bare
parameters adopted in the zero temperature runs: some entries refer
also to simulations already reported in Ref.~\cite{Bonati:2014ksa};
most of the new runs have been performed on the finest $48^3 \times
96$ lattice.

\begin{figure}[t!]
\centering
\includegraphics*[width=\columnwidth]{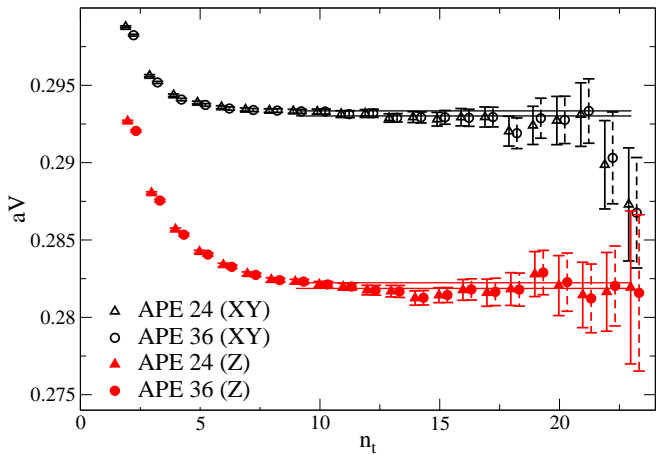}
\caption{Wilson loop combination $aV(a\vec{n}, an_t)$ defined in
  Eq.~\eqref{eq:potW} for $|\vec{n}|=3$ as a function of
  $n_t$. Results refer to two different values of the APE smearing
  level and to different orientations (orthogonal, $XY$, or parallel,
  $Z$) of the quark-antiquark separation relative to $\vec B = B \hat
  z$. The simulation has been performed on a $48^3 \times 96$ lattice
  at $|e|B\simeq 1\, \textnormal{GeV}^2$.}
\label{fig:wilsonloop}
\end{figure}

The Rational Hybrid Monte-Carlo (RHMC) \cite{Clark:2004cp,
  Clark:2006fx, Clark:2006wp} algorithm has been used to sample gauge
configurations. To determine the interquark potential in the confined
phase, statistics of $O(10^3)-O(10^4)$ trajectories 
have been collected for each value of the magnetic
background, with measures of Wilson loops performed every $5$
trajectories.

In order to reduce the UV noise, we used HYP \cite{Hasenfratz:2001hp}
and APE \cite{Albanese:1987ds} smearing in the following combination:
one step of HYP smearing for temporal links (with the parameters of
the HYP2-action reported in Ref.~\cite{Della Morte:2005yc}) and a
variable number of isotropic APE smearing steps (with parameter
$\alpha_{\mathrm{APE}}=0.25$), which has been chosen large enough to
significantly reduce the noise, but still small enough not to
introduce significant systematic effects. In practice, we verified
that a number of steps between $20$ and $40$ (depending on the lattice
spacing) satisfies these requirements. The potential was extracted
from planar Wilson loops of size $\vec{n}\times n_t$, making use of the
definition
\begin{equation}\label{eq:potW}
\begin{aligned}
& aV(a\vec{n}, an_t)\equiv \log\left(\frac{\langle W(\vec{n}, n_t )\rangle}{
\langle W(\vec{n}, n_t+1) \rangle}\right)\ , \\
& aV(a\vec{n})=\lim_{n_t\to\infty} aV(a\vec{n},n_t)\ ,
\end{aligned}
\end{equation}
i.e. by finding, for each fixed $\vec{n}$, a range of $n_t$ values
where the r.h.s. of the previous definition is stable, in order to
perform a fit to a constant. The fit range has been varied in order to
estimate the systematic error associated with its choice. It is
important to stress that since rotation symmetry is explictely broken
by the external field, different orientations of the Wilson loop have
to be studied independently in order to determine properly the
potential.

As an example of the procedure described above, in
Fig.~\ref{fig:wilsonloop} we report the values $aV(a\vec{n}, an_t)$ as
a function of $n_t$, obtained on the finest $48^3 \times 96$ lattice
with a magnetic background corresponding to $\vec b = 32\, \hat z$ ($|e|
B \sim$ 1 GeV$^2$). Data refer to the case $|\vec{n}|=3$ and we report
separately results obtained along the directions parallel or
transverse to $\vec B$, for two different APE smearing
levels.

For finite temperature simulations, the static quark-antiquark
potential has been determined from Polyakov loop correlators
\begin{equation}\label{eq:polycorr}
C(\vec{n},T)=\langle\mathrm{Tr}L(\vec{r}\,)\mathrm{Tr}\,
L^{\dagger}(\vec{r}+\vec{n}\,)\rangle
\end{equation}
where $\vec{r}$ and $\vec{n}$ are dimensionless lattice vectors. This
observable is related to the free energy $F_{Q\bar{Q}}(a \vec{n},T)$
of a static quark-antiquark pair separated by a distance $a \vec{n}$.
In principle, at a perturbative level, such correlator takes
contribution from both the singlet and octet color
channels~\cite{Brown:1979ya, McLerran:1981pb, Nadkarni:1986as,Kaczmarek:2002mc}, however one can show
that only the singlet contribution survives in the correlator in
Eq.~(\ref{eq:polycorr})~\cite{Jahn:2004qr,Rossi:2013qba}, so that one can
consistently define $C(\vec{n},T) \propto \exp( -F_{Q\bar Q} (a \vec
n, T)/T)$, where $F_{Q\bar Q}$ is the free energy of the static pair
in the singlet channel. Therefore, we shall adopt the definition
\begin{equation}\label{eq:freeenergy}
F_{Q\bar{Q}}(a \vec{n},T) = -\frac{1}{aN_t}\log C(\vec{n},T)
\end{equation}
which, apart from temperature dependent additive renormalizations, we
take as an estimate of the static quark-antiquark potential at finite
temperature ($N_t$ is the temporal size of the lattice, which is
related to the temperature of the system by $T=(aN_t)^{-1}$).  

For the finite temperature runs, we have adopted statistics comparable
to $T = 0$ runs (i.e. $O(10^3)-O(10^4)$ trajectories for each run),
with measures of the Polyakov loop correlators performed after each
trajectory and one step of HYP smearing in the temporal direction
(with the same parameters as for the $T = 0$ case) to suppress the UV
fluctuations.

\section{Results}
\label{results}

In this section we present the results of our simulations. We start with a
determination of the static potential at $T = 0$ and $B = 0$ for which, using
results at four different lattice spacings, we are able to obtain a reliable
continuum extrapolation.  We then move to the results obtained for $B \neq 0$:
we investigate the angular dependence of the anisotropic static potential, 
which 
is then continuum extrapolated using the numerical data for quark-antiquark
separations orthogonal or parallel to the external magnetic field.  We point
out some interesting features that seem to emerge in the limit of very large
magnetic field and finally we discuss the modifications induced by a
non-vanishing temperature.

\subsection{$T=0$, $B = 0$: continuum extrapolated results}\label{T0B0}

The zero temperature static quark-antiquark potential has been largely
investigated by means of phenomenological studies and lattice simulations and
it has been shown to be well described by the so-called Cornell potential
\cite{Eichten:1974af}  
\begin{equation}\label{eq:cornell_model}
V(r)=-\frac{\alpha}{r}+\sigma r + V_0\ ,
\end{equation}
where $\sigma$ is the string tension, $\alpha$ is the Coulomb parameter and
$V_0$ is an arbitrary constant term.  A potential related quantity that is
often convenient to introduce is the so-called Sommer parameter $r_0$, which is
defined by the equation \cite{Sommer:1993ce}
\begin{equation} \label{sommerdefinition}
r_0^2 \left. \frac{d V}{dr}\right|_{r_0} = 1.65 \, . 
\end{equation}
This parameter can be related to those entering 
Eq.~\eqref{eq:cornell_model} by
\begin{equation} \label{sommerrelation}
r_0 = \sqrt{\frac{1.65 - \alpha}{\sigma}} \, ,
\end{equation}
so that only two out of $r_0$, $\alpha$ and $\sigma$ are independent quantities.

\begin{figure}
\includegraphics*[width=\columnwidth]{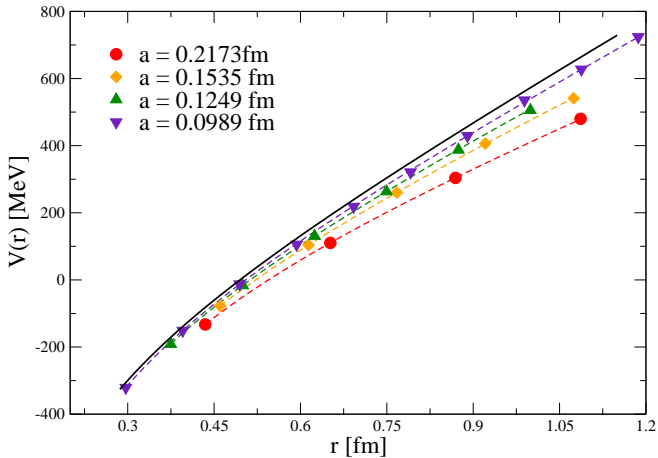}
\caption{Static potential as a function of the quark-antiquark distance in
physical units and for various lattice spacings. Dashed curves represent the
result of a fit according to the Cornell potential in
Eq.~(\ref{eq:cornell_model}), while the solid curve represents the continuum
extrapolation.  The constant $V_0$ has been shifted, for each lattice spacing
separately, so that the potential vanishes at $r_0$.}
\label{fig:pot_cont_b0}
\end{figure}

\begin{figure}[htb!]
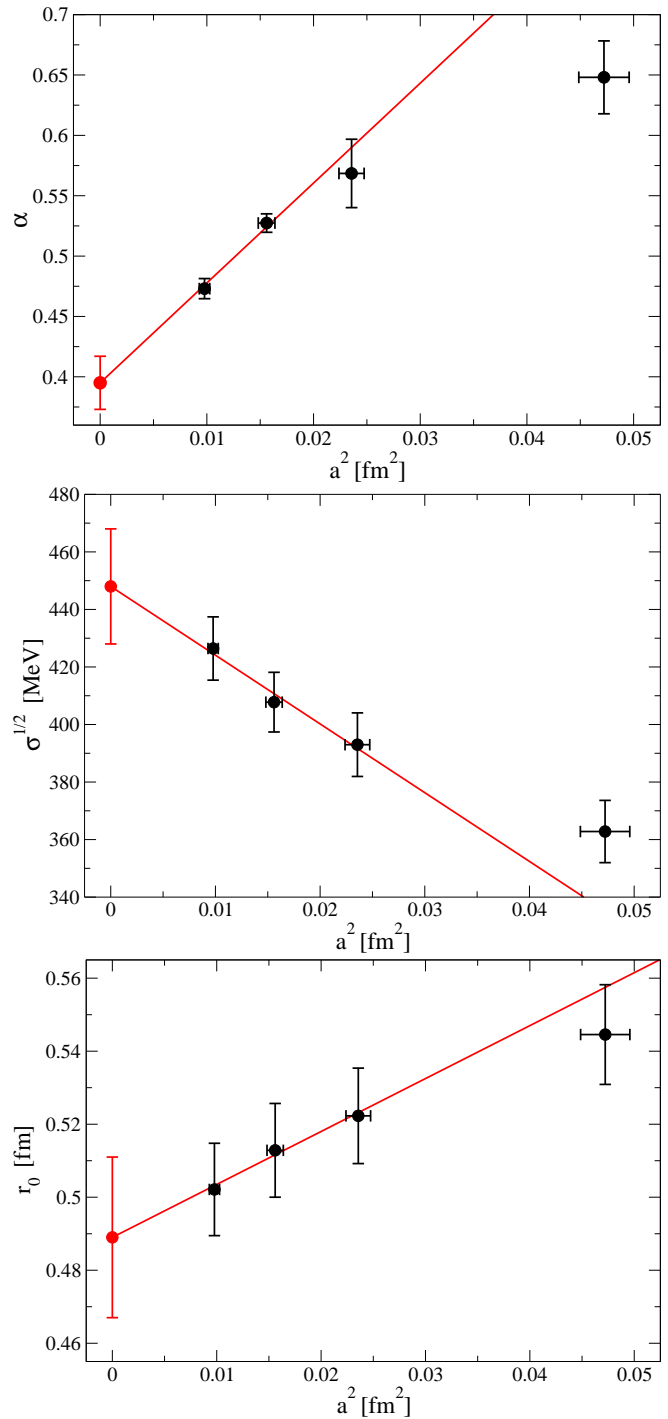

\centering
\includegraphics*[width=\columnwidth]{alpha_B0.eps}
\includegraphics*[width=\columnwidth]{sigma_B0.eps}
\includegraphics*[width=\columnwidth]{sommer_B0.eps}
\caption{Continuum extrapolation of the $Q\bar{Q}$ potential parameters
$\alpha,\sigma$ and $r_0$ at $T=0$ and $B = 0$.  The extrapolation has been
performed assuming $O(a^2)$ corrections for each parameter 
and considering only the three finest lattice spacings in each case, obtaining
respectively a reduced chi-squared $\chi^2/n_{dof} = 0.87/1$, 
$\chi^2/n_{dof}=0.12/1$ and
$\chi^2/n_{dof}=0.1/1$. Best-fit curves
are shown in the figures as well, together with the continuum extrapolated
values.}
\label{fig:continuum_B0}
\end{figure}

In Fig.~\ref{fig:pot_cont_b0} we show the results obtained for the potential,
using the procedure outlined in Section~\ref{numsetup}, for different lattice
spacings. In this case, since no magnetic background is present, Wilson loops
have been averaged over the different spatial directions.  For each
lattice spacing, data have been fitted according to 
Eq.~(\ref{eq:cornell_model}), in order to extract the values of $\sigma$,
$\alpha$ and $r_0$ ($\sigma$ and $r_0$ have been used as independent fit
parameters). In all cases the fit turned out to have a reduced
chi-squared around one and we verified the stability of the results against 
modifications of the fit range adopted to extract the parameters.

The values of $\sigma, r_0$ and $\alpha$ obtained at each lattice spacing will
be used in the following as reference values to determine the anisotropies
induced by the presence of the magnetic field. Their scalings with the
(square of the) lattice spacing are shown in Fig.~\ref{fig:continuum_B0}:
data are compatible with a linear dependence on $a^2$ in the whole explored
interval, however we have decided to keep only the three finest lattice
spacings to perform the continuum extrapolation.
Continuum values 
are reported in Table~\ref{tab:continuum_B0} 
(see also Fig.~\ref{fig:pot_cont_b0}), we have verified that 
they do not change appreciably, within errors, if data on the coarsest 
lattice are added to the fit and/or if a $O(a^4)$ term is added.

Values obtained for $r_0$ and $\sigma$ are in good agreement with
phenomenological estimates and with previous
lattice determinations~\cite{Blossier:2009bx,
Baron:2009wt, Fritzsch:2012wq, bali2013, Davies:2009tsa, Bazavov:2011nk,Aubin:2004wf}
(see also Section 9.2 of Ref.~\cite{Aoki:2016frl} for a recent review).

\begin{table}[t]
\centering
\begin{tabular}{|l|l|}
\hline
$\alpha$        & 0.395(22)\\
$\sqrt{\sigma}$ & 448(20) MeV\\
$r_{0}$         & 0.489(20) fm\\
\hline
\end{tabular}
\caption{Continuum extrapolated results for $\alpha,\sigma$ and $r_0$
(see also Fig.~\ref{fig:continuum_B0}).}
\label{tab:continuum_B0}
\end{table}

\subsection{$T=0$, $B \neq 0$: angular dependence}\label{angular}

The static potential studied in the previous subsection is isotropic, i.e. it
depends only on the modulus $r$ of the distance between the quark and 
the antiquark.
In Ref.~\cite{Bonati:2014ksa} it was shown that
this property is lost in the presence of a magnetic background which explicitly
breaks rotational invariance. The investigation carried out in
Ref.~\cite{Bonati:2014ksa} was restricted to the cases of quark-antiquark
separations parallel or orthogonal to the magnetic field, the final result of
that work being that in both these directions the potential is still well
described by Eq.~\eqref{eq:cornell_model}, but the parameters
depend on the direction.  In particular, the string tension is larger (smaller)
in the direction orthogonal (parallel) to $\vec B$, while $\alpha$ has 
an opposite behavior.

One of the purposes of our present investigation is to give a full description
of the static potential for $B~\neq~0$. There are some general features of the
angular dependence that can be fixed \emph{a priori}: while a generic anisotropic
potential is a function of the distance and of two angular variables, in the
case of a uniform magnetic field we can still rely on the residual rotational
symmetry around $\vec B$, thus reducing to a single angular variable
dependence, i.e. $V(r,\theta)$, where $\theta$ is the angle between the
quark-antiquark separation and $\vec B$. Furthermore, since one expects in
general symmetry under inversion of $\vec B$, we require $V(r,\theta) = V(r,\pi
- \theta)$.  Motivated by the results of Ref.~\cite{Bonati:2014ksa} we make the 
ansatz that for each fixed value of $\theta$ the Cornell description still
holds, i.e. that 
\begin{equation}\label{angular_cornell} 
V(r,\theta,B) = \sigma (\theta,B)\, r - \frac{\alpha(\theta,B)}{r} + V_0(\theta,B) \, .
\end{equation}
The validity of such an assumption for all values of $\theta$ can of course
only be verified \emph{a posteriori}.  Notice that we assume that also the 
constant
term $V_0$ in the Cornell potential can take an angular dependence.

The most general description of the angular dependence of the 
Cornell parameters can be given in terms of a Fourier series,
for instance for the string tension we can write 
\begin{equation}
\sigma(\theta,B) = \bar \sigma (B) 
\left( 1 - \sum_{n=1}c^{\sigma}_{2n}(B) \cos (2n\theta)\right)
\end{equation}
where only Fourier coefficients which respect the symmetry under $\theta \to
\pi - \theta$ have been used. Our general parametrization will therefore be the
following one:
\begin{equation}\label{eq:cornell_angular}
\begin{aligned}
V(r,\theta) = &-\frac{\bar\alpha(B)}{r}\Big(1-\sum_{n=1}c^{\alpha}_{2n} (B)\cos(2n\theta)\Big) \\
&+\bar\sigma(B) r \Big(1-\sum_{n=1}c^{\sigma}_{2n} (B) \cos (2n\theta)\Big) \\
&+\bar V_0 (B) \Big(1-\sum_{n=1}c^{V_0}_{2n} (B) \cos(2n\theta)\Big) \, .
\end{aligned}
\end{equation}
How many Fourier coefficients are required to reliably describe the actual
potential can be decided only on the basis of numerical
results, that we are going to expose and discuss in the following.

In order to study the complete angular dependence of the potential one
could in principle proceed in two different ways: by rotating either
the Wilson loop or the magnetic field.  In the first case a rotation
of the spatial side of the Wilson loop by small angles (i.e. by less
than $\pi/2$), would require a significant modification of the spatial
path; in particular new cusps appear, which significantly modify the
renormalization factors and make the comparison of results at
different angles and lattice spacings more involved.  For this reason
we choose to rotate the magnetic field, a choice that however requires
to perform new Monte Carlo simulations for different orientations of
the magnetic field, since $\vec{B}$ enters directly into the
probability distribution of gauge configurations. As a consequence we
fully investigated just a limited set of angular orientations, that
however, as we will show, is sufficient to make the picture clear
enough.

In simulations with $\vec{B}$ not directed along any of the lattice
axes, in the computation of ratios appearing in Eq.~\eqref{eq:potW} we
have considered separately the Wilson loop directed along the $X$, $Y$
and $Z$ directions, which in general correspond to three different
values of $\theta$.  Simulations have been performed for two lattice
spacings only, namely $a=0.0989$ and $0.1535$ fm, using lattices of
size $48^3 \times 96$ and $32^4$ respectively.  In both cases we
considered a single value of the modulus of the magnetic field,
$|\vec{b}|\simeq 32$, corresponding $|e|B\sim 1\,\mathrm{GeV}^2$,
since the spatial physical size is consistently $a L_s\simeq
5\,\mathrm{fm}$. We studied the following combinations of magnetic
field quanta: $(b_x,b_y,b_z)=(0,0,32)$, $(4,13,29)$ and $(9,18,25)$,
which give access to $8$ different values of the angle $\theta$.

\begin{figure}
\centering
\includegraphics*[width=\columnwidth]{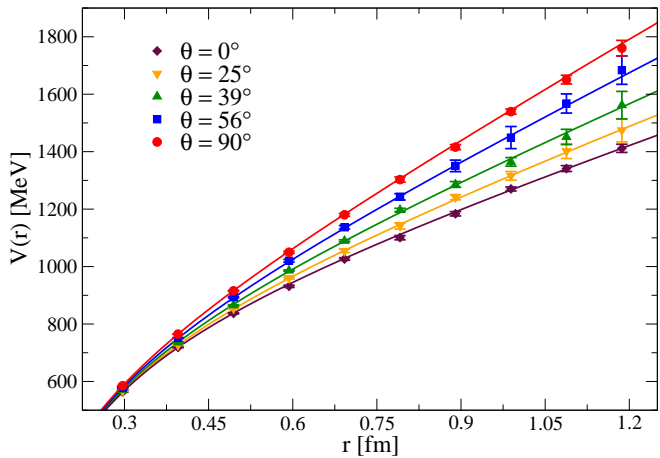}
\caption{The potential obtained from the $48^3\times96$ lattice, for several
values of the angle $\theta$ between the Wilson loop and $\vec{B}$ with
$|e|B\simeq1$ GeV$^2$. Best fit curves are obtained using the form in
Eq.~\eqref{eq:cornell_angular}.}
\label{fig:potential_angles}
\end{figure}

In Fig.~\ref{fig:potential_angles} we show the results obtained for the 
potential
as a function of $r$ for different values of $\theta$ on the finest, $48^3
\times 96$ lattice. A property of its angular dependence is clearly visible,
which is present also for data on the coarser $32^4$ lattice and is in line
with what observed in Ref.~\cite{Bonati:2014ksa}: at fixed $r$, the potential
increases as the angle between the quark-antiquark separation and $\vec B$
increases, and reaches its maximum in the plane orthogonal to $\vec B$.

\begin{figure}
\centering
\includegraphics*[width=\columnwidth]{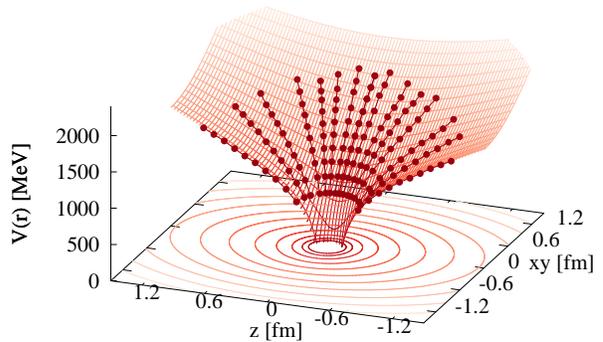}
\caption{3D-plot of the potential as a function of the spatial coordinates. Data points refer to the
$48^3\times96$ lattice at $|e|B\sim1$ GeV$^2$ (with $\vec{B}$ directed along
$Z$) and are fitted by the surface defined in Eq.~\eqref{eq:cornell_angular}.}
\label{fig:potential_3d}
\end{figure}

\begin{figure}
\centering
\includegraphics*[width=1.15\columnwidth,height=0.75\columnwidth]{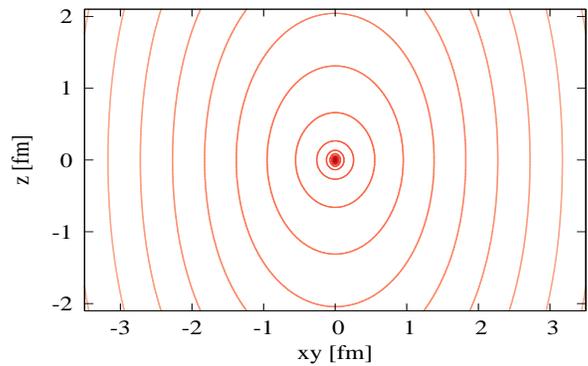}
\caption{Contour plot of the function displayed in Fig.~\ref{fig:potential_3d}. 
Contour lines are plotted every 500 MeV.}
\label{fig:potential_contour}
\end{figure}

\begin{figure}[t!]
\centering
\includegraphics*[width=\columnwidth]{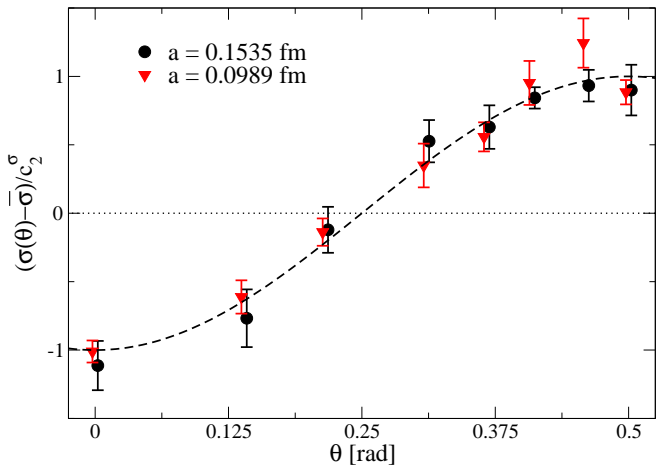}
\caption{Rescaled form of the string tension as function of the angle $\theta$
for the two lattices $32^4$ and $48^3\times96$. The dashed line is the function
$-\cos(2\theta)$.}
\label{fig:sigma_angular_dependence}
\end{figure}

A peculiar property that we found is that, as one tries to perform a 
best fit of data according to 
the ansatz given in Eq.~(\ref{eq:cornell_angular}),
it is sufficient to include only the lowest order Fourier coefficient $c_2$,
for each parameter, in order to obtain a reasonable value of the reduced
chi-squared test (see the caption of Table~\ref{tab:potential_angular_fit}),
and if further coefficients are inserted in the best fit, they
come out to be compatible with zero within errors.
The best fit function obtained for the finer lattice is displayed in
Fig.~\ref{fig:potential_angles} as a function of $r$ and $\theta$,
while in Fig.~\ref{fig:potential_3d} the same best fit function is
shown in a three dimensional plot. The anisotropy of the potential is
better seen by looking at the contour plot in
Fig.~\ref{fig:potential_contour}.  The best fit parameters for both
lattices are reported in Table~\ref{tab:potential_angular_fit}, and
in Fig.~\ref{fig:sigma_angular_dependence} we show the angular
variation of the string tension, from which it is clear that a
single cosine term well describes data in both cases.

\begin{table}[htb!]
\centering
\begin{tabular}{|c|ccc|}
\hline
\rule{0mm}{3.5mm}   & $\bar{\alpha}$ & $\sqrt{\bar{\sigma}}$ [MeV] & $\bar{V}_0$ [MeV]\\
$48^3\times96$      & 0.493(6)       & 414(2)                      & 644(5)\\
$32^4$              & 0.499(23)      & 398(4)                      & 407(20)\\
\hline
\end{tabular}\\
\begin{tabular}{|c|ccc|}
\hline
{}             & $c^{\alpha}_2$ & $c^{\sigma}_2$ & $c^{V_0}_2$\\
$48^3\times96$ & -0.130(10)     & 0.262(7)       & -0.154(8)\\
$32^4$         & -0.323(64)     & 0.351(32)      & -0.428(52)\\
\hline
\end{tabular}
\caption{Best fit parameters for the potential in
Eq.~\eqref{eq:cornell_angular}. The reduced chi-square values are
$\chi^2/n_{dof} = 53/74$ and $\chi^2/n_{dof} =36/34$ for the
$48^3\times96$ and the $32^4$ lattice respectively. 
Note that in this particular case the value of the chi-square is 
only a qualitative 
estimator of the fit goodness, since data are correlated (correlations have been 
propagated using a bootstrap procedure).
}
\label{tab:potential_angular_fit}
\end{table}

Another interesting feature emerges from the data in
Table~\ref{tab:potential_angular_fit}: the constant term in the
Cornell potential gets an angular dependence as well. This is a very
peculiar feature, since that would imply an additive term in the potential
which depends on $\theta$ but not on $r$. However, we stress that the
associated Fourier coefficient $c_2^{V_0}$ is reduced by a factor
$\sim 2.8$ as one moves from the coarse to the fine lattice,
consistently with the vanishing of such angular dependence in the
continuum limit. Notice that an analogous consideration, i.e. the
vanishing of the angular dependence in the continuum limit, could be
made for the Coulomb coupling, but not for the string tension; a more
detailed discussion of this issue will be done in the following
subsection.  Finally, we stress that most of the $B$-dependence of the
potential can be ascribed to Fourier coefficients, since the
parameters $\bar \alpha$, $\bar \sigma$ and $\bar V_0$, which are sort of
averaged values over different directions, show just a mild
variation with respect to the $B = 0$ case.

Having determined that the full angular dependence of the static quark
potential can be described just by the first non trivial harmonic
term, we can determine all relevant parameters from an analysis of the
potential along the directions parallel and orthogonal to $\vec B$,
i.e. from $\theta = 0, \pi/2$, which are the standard quantities
already explored in Ref.~\cite{Bonati:2014ksa} and which have been
extended in the present study to a finer lattice spacing and to
different values of $B$.

\subsection{$T=0$, $B \neq 0$: continuum extrapolated results}\label{sec:continuum_anisotropy}

In this subsection, in order to perform a continuum extrapolation of
the $B$-dependence of the static potential, we consider numerical
results obtained at different values of the magnetic field and of the
lattice spacing, and mostly for quark-antiquark separation parallel
($Z$) or orthogonal ($XY$) to the magnetic field.  That gives us
access to the Cornell parameters in those directions,
i.e. $\alpha_{XY}$ and $\alpha_{Z}$, $\sigma_{XY}$ and $\sigma_{Z}$,
$V_{0,XY}$ and $V_{0,Z}$, $r_{0,XY}$ and $r_{0,Z}$.  According to the
analysis of the angular dependence given in the previous subsection,
that provides us with enough information to characterize the complete
behavior of the potential. Indeed, if for each parameter $\mathcal{O}$
we introduce its anisotropy
\begin{equation}\label{eq:anisotropy}
\delta^{\mathcal{O}}(|e|B)=
\frac{\mathcal{O}_{XY}(|e|B)-\mathcal{O}_{Z}(|e|B)}{\mathcal{O}_{XY}(|e|B)+\mathcal{O}_{Z}(|e|B)}
\end{equation}
and its average relative change with $B$
\begin{equation}
R^{\mathcal{O}}(|e|B)=
\frac{\mathcal{O}_{XY}(|e|B)+\mathcal{O}_{Z}(|e|B)}{2 \mathcal{O} (|e|B = 0)}
\end{equation}
we have, assuming that $c_{2n}^{\mathcal{O}} \simeq 0$ when $n > 1$, 
\begin{equation}\label{eq:anisotropy_coefficients}
\delta^{\mathcal{O}}=c_2^{\mathcal{O}}+c_4^{\mathcal{O}}+\dots=
\sum_{n}c_{2n}^{\mathcal{O}} \simeq c_2^{\mathcal{O}}
\end{equation}
and
\begin{equation}\label{ratiodef}
\begin{aligned}
R^{\mathcal{O}}(|e|B) &= 
\frac{\bar {\mathcal{O}} (|e| B)}{\mathcal{O} (|e| B = 0)} \left( 
1  -\sum_{n\ {\rm even}} c_{2n}^{\mathcal{O}} \right)  \\
&\simeq \frac{\bar {\mathcal{O}} (|e| B)}{\mathcal{O} (|e| B = 0)}
\end{aligned}
\end{equation}
i.e. such quantities are enough to fix all the coefficients
giving a non-trivial contribution to Eq.~(\ref{eq:cornell_angular}).
That would have not been possible without the assumption
$c_{2n} \simeq 0$ for $n > 1$.

\begin{figure}[!htb]
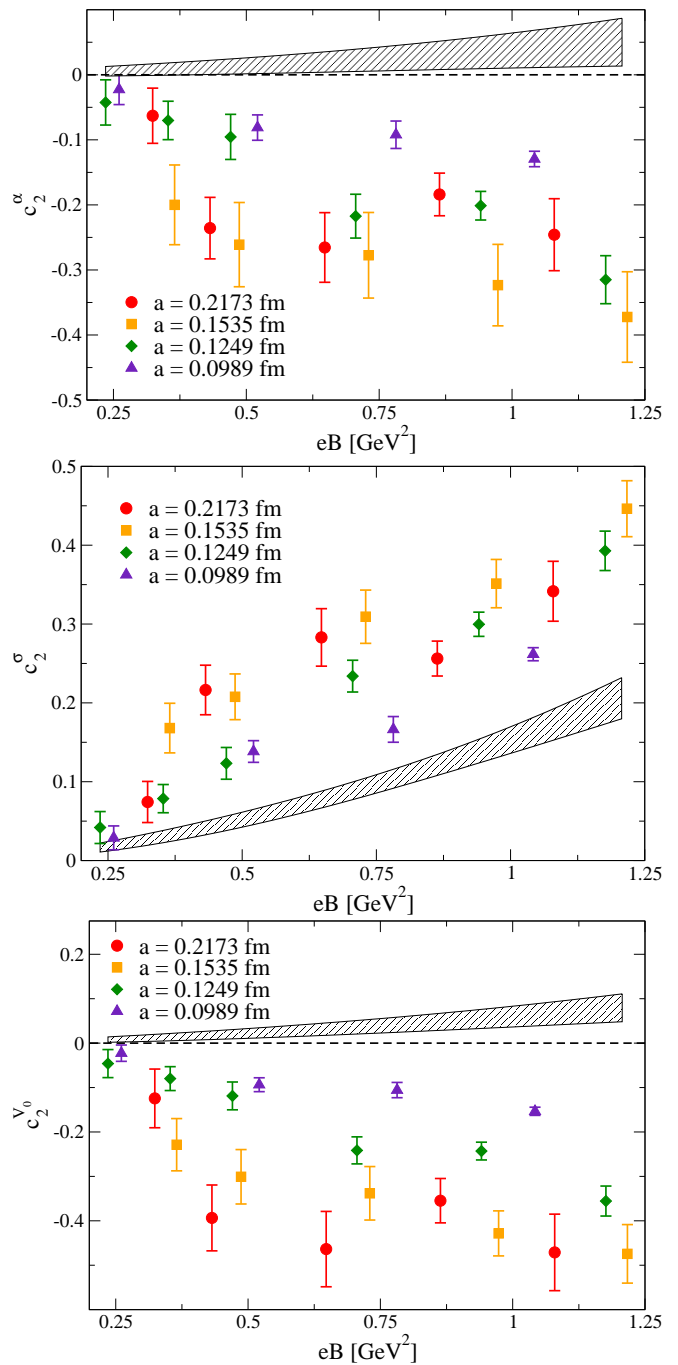

\centering
\includegraphics*[width=\columnwidth]{alpha_aniso.eps}
\includegraphics*[width=\columnwidth]{sigma_aniso.eps}
\includegraphics*[width=\columnwidth]{consta_aniso.eps}
\caption{The anisotropy coefficients $c^{\mathcal{O}}_2$ for all the
  potential parameters $\mathcal{O}=\alpha,\sigma,V_0$. Data 
have been computed according to the definition in
  Eq.~\eqref{eq:anisotropy}. Gray bands represent the continuum
  extrapolation obtained by fitting with Eq.~\eqref{eq:ratios_ansatz}
  all data except the ones of the coarsest lattice $24^4$.}
\label{fig:anisotropy}
\end{figure}

\begin{figure}[!htb]
\centering
\includegraphics*[width=\columnwidth]{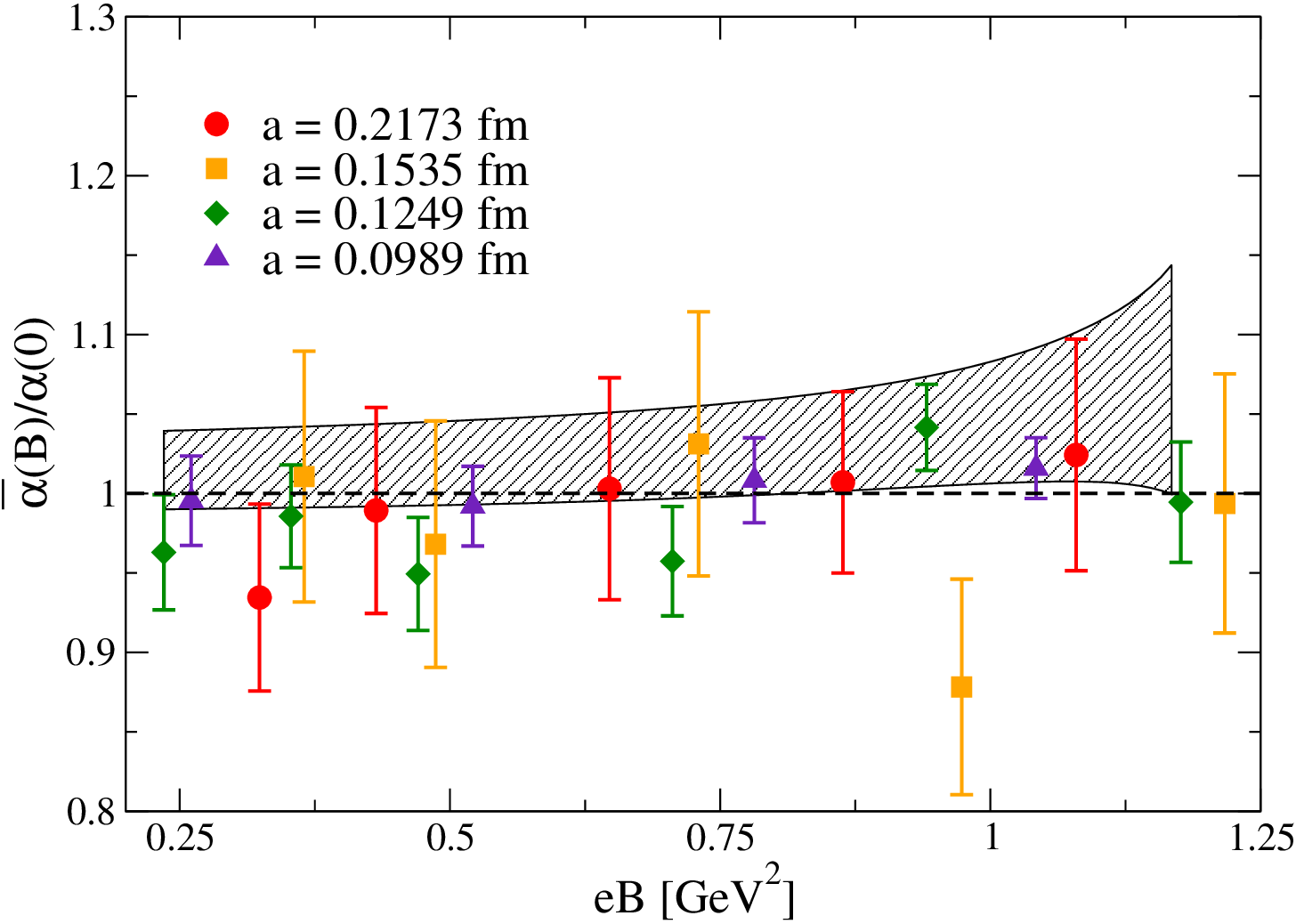}
\includegraphics*[width=\columnwidth]{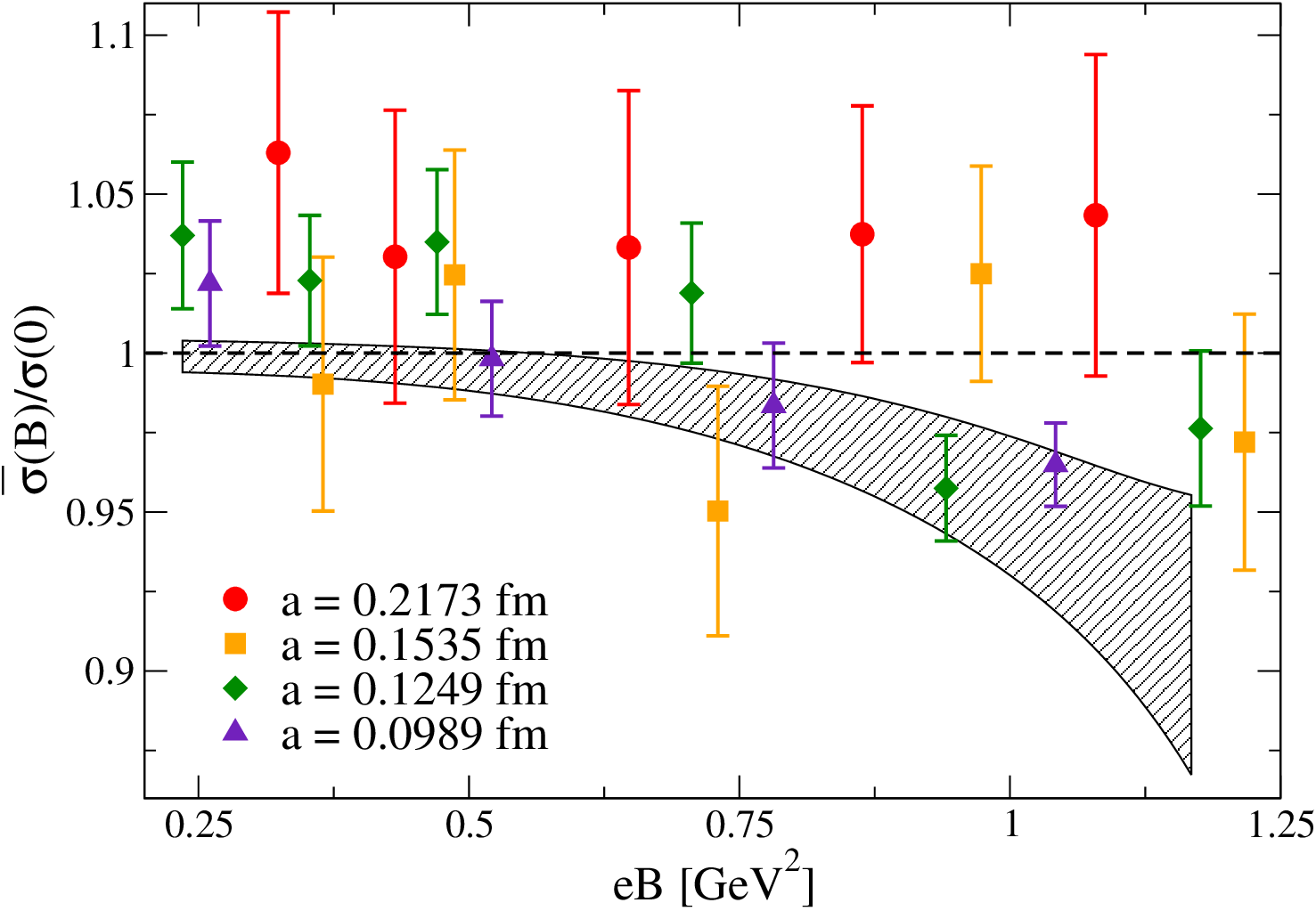}
\includegraphics*[width=\columnwidth]{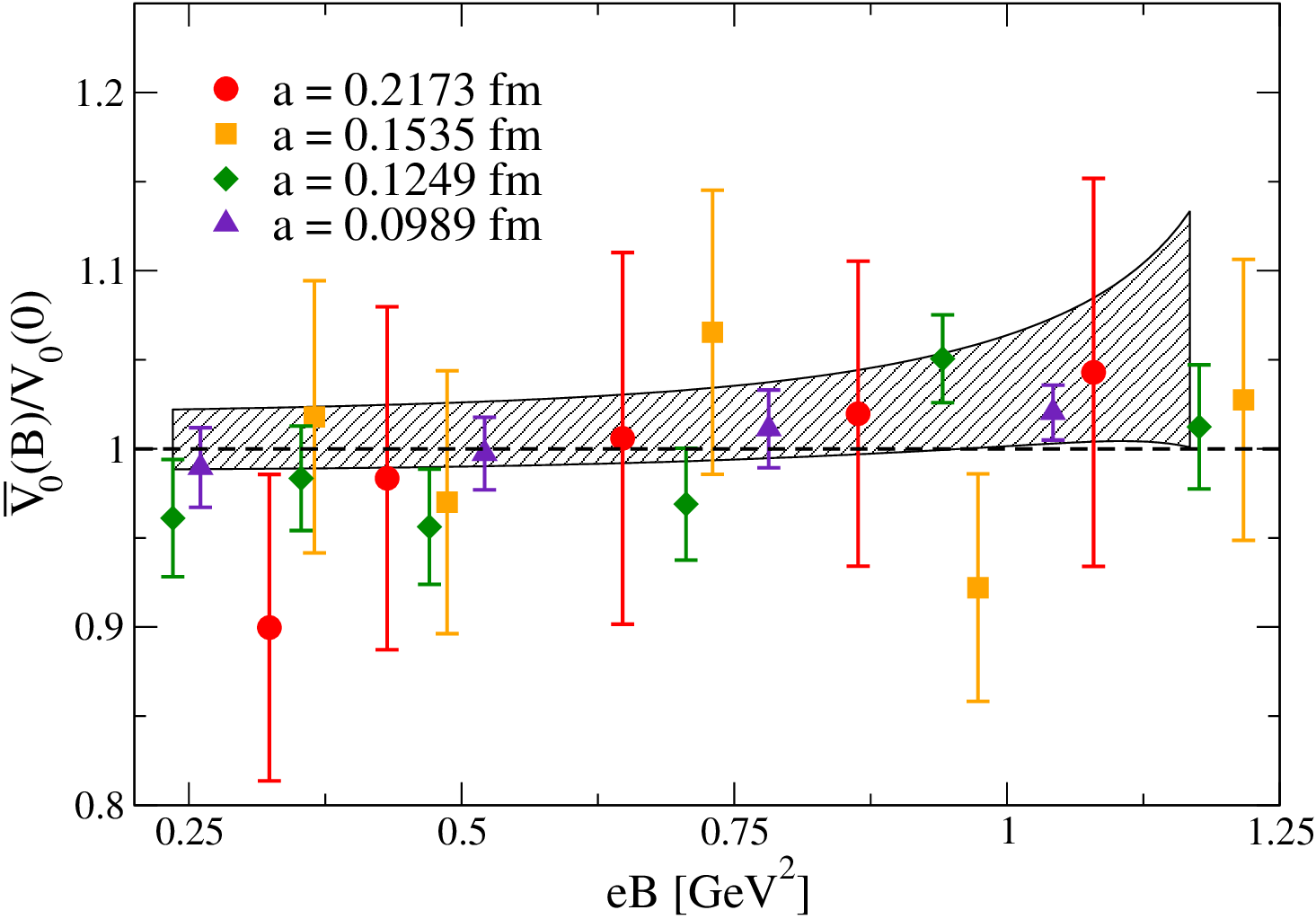}
\caption{The ratios $R^\mathcal{O}$ for all the potential parameters
$\mathcal{O}=\alpha,\sigma,V_0$. Data 
have been computed
according to the definition in Eq.~\eqref{ratiodef}. Gray bands represent the
continuum extrapolation obtained by fitting with Eq.~\eqref{eq:ratios_ansatz}
all data except the ones of the coarsest lattice $24^4$.}
\label{fig:all_ratios}
\end{figure}

Results obtained in this way for the $c_2$ coefficients at the various
magnetic fields and lattice spacings are shown in
Fig.~\ref{fig:anisotropy}, while in Fig.~\ref{fig:all_ratios} we
report results for the quantities $R^{\mathcal{O}}(|e|B)$. A
remarkable feature is that this quantities show a very mild variation
with the magnetic field, so that most of the dependence of the
potential on $B$ can be ascribed to the anisotropy coefficients $c_2$.
Moreover, the signs of such anisotropies are always consistent with
the fact that, at fixed $r$, the potential has a minimum (maximum) for
$\theta = 0 \,(\pi/2)$.

We performed a continuum limit extrapolation of our data according to
the following ansatz
\begin{equation}\label{eq:ratios_ansatz}
\begin{aligned}
& c_2^{\mathcal{O}}= A^{\mathcal{O}}(1+C^{\mathcal{O}}a^2)(|e|B)^{D^{\mathcal{O}}(1+E^{\mathcal{O}}a^2)} \\
& R^{\mathcal{O}}=1+\bar A^{\mathcal{O}}(1+ 
  \bar C^{\mathcal{O}}a^2)(|e|B)^{\bar D^{\mathcal{O}}}
\end{aligned}
\end{equation}
which is similar to the one adopted in Ref.~\cite{Bonati:2014ksa} and
consists of a power law in $B$ for both set of quantities, with the
insertion of $O(a^2)$ corrections for all involved
coefficients. Reasonable fits are obtained if the data on the coarsest
lattice are discarded.  Continuum extrapolated results are displayed
in Fig.~\ref{fig:anisotropy} and Fig.~\ref{fig:all_ratios}, while
numerical results for the best fit parameters are reported in
Table~\ref{tab:anisotropy_results}.

The results obtained for the $c_2$ coefficients show that, while the
anisotropy of the string tension has a well defined non-zero continuum
limit, with a value around 15~\% for $|e| B \sim 1$ GeV$^2$, those of
the Coulomb coupling and of $V_0$ seem to disappear in the same
limit. Indeed, reasonable best fits are obtained also by 
imposing $c_2^\alpha$
and $c_2^{V_0}$ to vanish in the continuum limit
($\chi^2/n_{dof} \sim 9.4/12$ and $\chi^2/n_{dof} \simeq 14.4/12$
respectively for $c_2^\alpha$ and $c_2^{V_0}$), while the same is not true for
$c_2^{\sigma}$ ($\chi^2/n_{dof} \simeq 87/12$).
The continuum
extrapolated results obtained for the $R^{\mathcal{O}}$ ratios
confirm, instead, that on average all Cornell parameters have little
dependence on the magnetic field. 
Again, reasonable best fits are obtained also by 
imposing $R^{\mathcal{O}}$ 
to be $B$-independent 
in the continuum limit ($\chi^2/n_{dof} = 8.3/12 $ for 
$\bar{\alpha}$, $\chi^2/n_{dof} = 9.1/12 $ for $\bar{V_0}$, and $\chi^2/n_{dof} = 
18/12$ for $\bar{\sigma}$).

\begin{figure}[htb!]
\centering
\includegraphics*[width=\columnwidth]{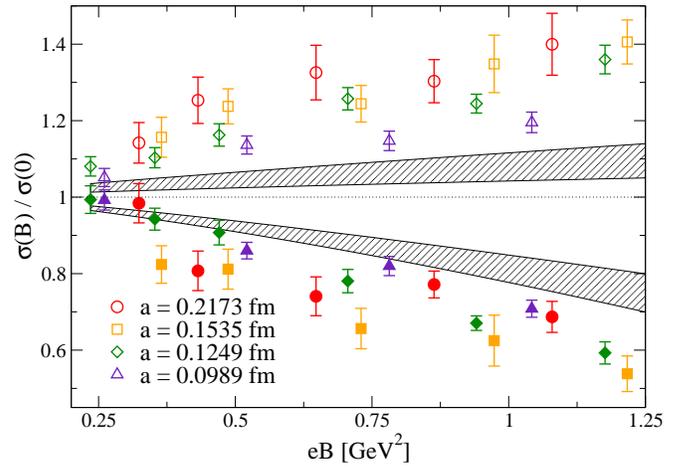}
\caption{Ratios between the string tension computed along the $XY$ (empty symbols) and 
the $Z$ (full ones)
directions and the string tension computed at $B=0$.  Bands denote the continuum
extrapolations, that have been obtained using an ansatz analogous to the
one used in Eq.~(\ref{eq:ratios_ansatz}) on the three finest lattice spacings.
Best fit parameters (relevant in the continuum limit) are $A^{\sigma_{XY}} = 
0.084(42) $, $D^{\sigma_{XY}} =0.90(37)$, with $\chi^2/n_{dof} = 8.9/11 $, and 
$A^{\sigma_{Z}} = -0.198(39) $, $D^{\sigma_{Z}} =  2.08(44)$ with 
$\chi^2=10.6/11 $.}
\label{fig:ratio}
\end{figure}

In summary, the modification of the quark-antiquark potential induced by the
magnetic field persists in the continuum limit, and this is mostly due to the
anisotropy induced in the string tension. In order to directly compare with
the analysis of Ref.~\cite{Bonati:2014ksa}, in Fig.~\ref{fig:ratio} we report
also results for the relative variation of the string tension along the $XY$
and $Z$ axis, together with their continuum extrapolations.

\begin{table}[htb!]
\centering
\begin{tabular}{ |c|c||c|c||c|c| }
\hline
$A^{\sigma}$ & 0.151(32) & $D^{\sigma}$ & 1.64(30) & 
$\chi^2/n_{dof}$ & 9.5/11\\
$A^{\alpha}$ &0.046(39) & $D^{\alpha}$ & 1.51(67) & 
$\chi^2/n_{dof}$ & 7.3/11\\
$A^{V_0}$ & 0.066(36) & $D^{V_0}$ & 1.48(50) & 
$\chi^2/n_{dof}$ & 7.8/11\\
\hline
\end{tabular}
\caption{Continuum limit of the anisotropies defined in
Eq.~\eqref{eq:anisotropy}, performed using the ansatz in
Eq.~\eqref{eq:ratios_ansatz}. The fit does not include the coarse $24^4$
lattice.}
\label{tab:anisotropy_results}
\end{table}

\subsection{Anisotropic deconfinement in the large field limit?}

\begin{figure}[t!]
\centering
\includegraphics*[width=\columnwidth]{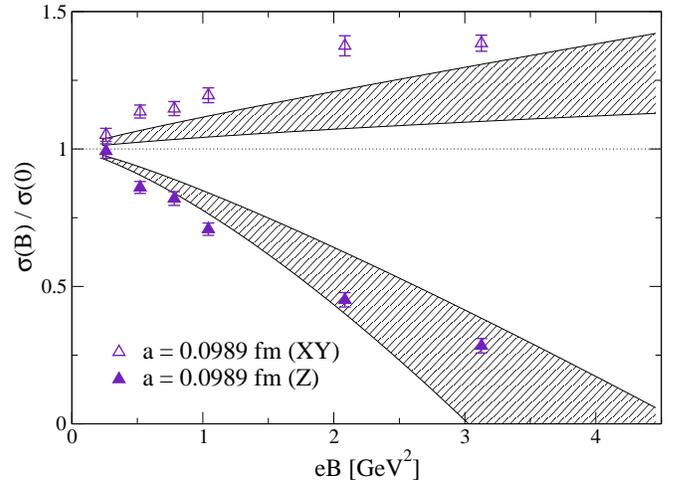}
\caption{Relative change of the string tension as a function of $|e| B$,
along the directions parallel and orthogonal to the background field. 
The shadowed band corresponds to the continuum extrapolation, which however
is based on results obtained for $|e| B \lesssim 1$ GeV$^2$.}
\label{fig:largeB}
\end{figure}

The continuum extrapolation discussed in the previous subsection has
been obtained for a range of magnetic fields going up to around 1
GeV$^2$, however it is tempting to extrapolate it to larger values of
$B$. In particular, results obtained for the longitudinal string
tension indicate that it would vanish for magnetic fields of the order
of 4 GeV$^2$: this is clearly visible from Fig.~\ref{fig:largeB},
where the continuum extrapolation for the ratio
$\sigma(|e|B)/\sigma(0)$ along the axes has been extended up to
$eB\sim 4\mathrm \, {\rm GeV}^2$.

A vanishing string tension at some critical value of $B$,
corresponding to a sort of longitudinal deconfinement, would have
important consequences, e.g. for the anisotropic propagation of heavy
quark-antiquark pairs, and could be put in connection with theoretical
speculations about anisotropic quantum transitions at large values of
$B$~\cite{chernosuper}.

A question naturally arises at this point: how reliable is the
extension to large magnetic fields of a continuum extrapolation
obtained in a smaller range of $B$? Unfortunately, a direct continuum
extrapolation for $|e| B \gg 1$ GeV$^2$ is presently hindered by the
large ultraviolet cut-off effects which are expected when $|e| B
\gtrsim 1/a^2$~\cite{review}. This sets the limit $|e| B \lesssim 1$
GeV$^2$ for two of the lattice spacings explored in the present study.

However, we can extend the range of explored $|e| B$ just for the
finest lattice spacing, $a \simeq 0.0989$ fm, where $1/a^2 \sim 4$
GeV$^2$. Therefore, we have performed further numerical simulations
for $|e| B \sim 2$ and $3$ GeV$^2$, obtaining the results for the
string tension which are reported in Fig.~\ref{fig:largeB}. The new
results follow roughly the extrapolation of the 
continuum band, thus proving that
a steady increase of the anisotropy persists up to very large fields.
On the other hand, understanding whether an anisotropic deconfinement
really takes place at some critical value of $B$ requires further
studies at finer values of the lattice spacing.

\subsection{$T>0$ results}

\begin{figure}[h]
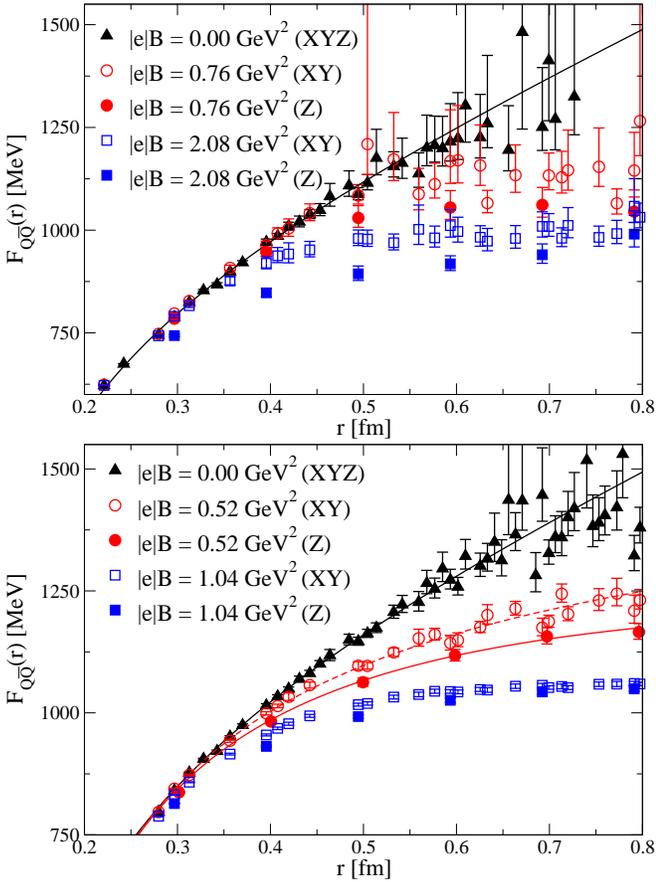

\includegraphics*[width=\columnwidth]{corr_finiteT_nt20.eps}
\includegraphics*[width=\columnwidth]{corr_finiteT.eps}
\caption{\emph{Top}: Free energy of  the $Q\bar{Q}$ pair $F_{Q\bar{Q}}(r,T)$ at 
$T~\simeq~100$~MeV ($48^3 \times 20$ lattice)  as a function of 
the distance $r$ for several values of $B$. \emph{Bottom}: $F_{Q\bar{Q}}(r,T)$  
at $T~\simeq~125$~MeV ($48^3\times 16$ lattice).
Continuum and dashed 
curves correspond to best fits to a Cornell potential, and are
reported only for cases in which the best fit works well in a reasonable 
range of distances.}
\label{fig:pot_finiteT}
\end{figure}

For finite temperature runs, we have extracted the 
static potential from the gauge invariant correlator
of Polyakov loops, see Eqs.~(\ref{eq:polycorr}) and
(\ref{eq:freeenergy}). In this case we report on results
at a single value of the lattice spacing 
(the finest one, $a \simeq 0.0989$ fm) and for
three temperatures below the transition temperature,
namely $T \simeq 100, 125, 143$ MeV (corresponding to lattices
$48^3 \times 20,16,14$, respectively), where we still expect
the system may exhibit confining properties.
In the deconfined phase, instead, Polyakov loop correlators give
access to various kind of gluonic screening masses: the effect of the
magnetic field on those masses will be studied in a forthcoming
investigation. In Fig.~\ref{fig:pot_finiteT} we show the results
obtained for $F_{Q \bar Q}$ at some values of the magnetic field,
respectively for $T~\simeq~100$ MeV and $T~\simeq~125$~MeV.  In
general, we observe a behavior which is different from that present at
$T = 0$.

The anisotropy is still visible, and goes
in the same direction as for $T = 0$, with the potential 
for quark-antiquark separations parallel
to $\vec B$ being suppressed with respect to the orthogonal case.
The anisotropy is more pronounced for large magnetic fields and for
intermediate separations, while it disappears in the limit of large
distances: this is actually expected, since the magnetic
background cannot disrupt the cluster property of the
theory, hence the correlation of two Polyakov loops must be independent, in
the large distance limit, from the direction of their relative 
separation.

\begin{figure}[h]
\includegraphics*[width=\columnwidth]{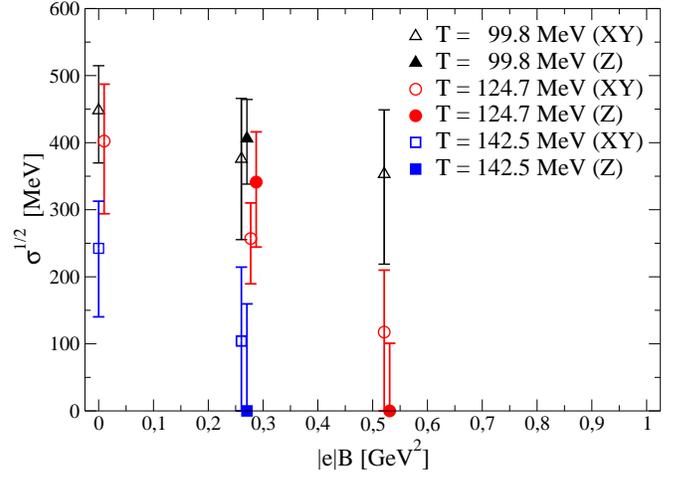}
\caption{The square root of the string tension $\sigma$ along different 
directions is reported, 
as a  function of $|e|B$, for three different temperatures below
  $T_c$ and for $a = 0.0989$ fm. 
In the case $|e| B = 0$ data refer to an average over all directions.}
\label{fig:stVSb}
\end{figure}

\begin{figure}[h]
\includegraphics*[width=\columnwidth]{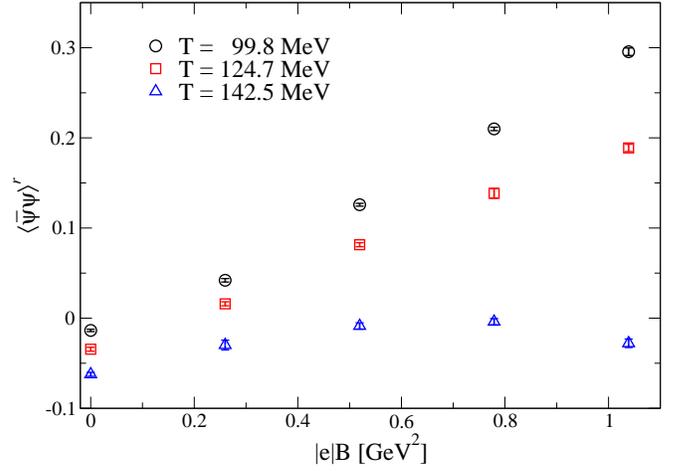}
\caption{Renormalized chiral condensate as a function of 
the magnetic field for various temperatures.}
\label{fig:chiral}
\end{figure}

Contrary to the zero temperature case, the main effect of the magnetic
field seems to be that of suppressing the potential in
all directions. Only for relatively small magnetic fields (smaller and
smaller as the temperature increases) the observed $F_{Q \bar Q}$ can
be fitted according to a Cornell potential and is qualitatively
similar to that observed for $T = 0$. However, in these cases one is
not able to distinguish, within errors, the results obtained in the
various directions: this is due to the larger statistical
uncertainties affecting the Polyakov loop correlator, especially in
the low temperature regime, where the Polyakov loop is more suppressed
and the correlators are noisier.  The results obtained for the
string tension in such cases are reported in Fig.~\ref{fig:stVSb}. As
expected for vanishing $|e|B$, the string tension $\sigma$ decreases
as the temperature approaches the deconfinement transition. Such an
effect comes out to be enhanced when the external field is turned on.

The observed behavior is consistent with the decrease of
the chiral pseudocritical temperature which has been observed
in previous studies~\cite{Bali:2011qj,Bali:2012zg}. 
This phenomenon has been usually named 
as {\em inverse magnetic catalysis}, because of its
relation with the behavior of the chiral condensate, which is
an increasing function of $B$ at low $T$ (direct magnetic catalysis),
and becomes non-monotonic around the transition, in association
with a decrease of $T_c$. Our observations provide evidence for 
a strong effect of the magnetic field also on  the confining properties
of the medium which, at fixed temperature, seem to be lost if the magnetic
field is strong enough. 

Actually, looking at the behavior of the chiral condensate 
for the same values of temperature and magnetic field, 
the effect on the confining properties seems to be the leading
phenomenon. In Fig.~\ref{fig:chiral} we report
the renormalized light chiral condensate as a function of $B$ for 
the three explored temperatures, which is defined as~\cite{Bali:2011qj}:
\beq
\langle \bar{\psi}\psi\rangle^r (T,B)=\frac{m_{l}}{m_{\pi}^4}\left(\langle\bar{\psi}\psi\rangle_{l} (T,B)
-\langle\bar{\psi}\psi\rangle_{l}(0,0)\right)\, 
\eeq
where the $T = 0$ subtraction eliminates additive divergences, while
the multiplication by the bare light quark mass $m_l$ 
takes care of multiplicative ones.
It is clearly visible that, for the two lowest temperatures, no signal
of inverse magnetic catalysis is visible in the explored range 
of $B$, while in the same range we already observe a strong
modification of the free energy of the static pair, 
such that we are not able to fit it according to a Cornell potential,
something which we interpret as an effective disappearance of the 
confining properties.

This might have many interesting interpretations and consequences.
First, from a theoretical point of view, the suppression of the
confining properties seems to be a dominant phenomenon with respect to
the effect of the magnetic field on the chiral condensate, so that one
would be tempted to describe the observed decrease of $T_c$ with $B$
in terms of ``deconfinement catalysis'' rather than of ``inverse
magnetic catalysis''; notice that this point of view
is also suggested by recent computations in holographic 
models~\cite{Dudal:2015wfn}. 
Second, from a phenomenological point of view,
the precocious modification of the confining properties induced by the
magnetic background, might have significant consequences on the
suppression of heavy quark bound states (e.g., $J/\psi$) 
to be observed in the thermal medium produced in non-central
heavy ion collisions.

\section{Conclusions}
\label{concl}

Following an explorative study~\cite{Bonati:2014ksa},
which showed the presence of strong effects on the confining 
properties of the QCD vacuum induced by the presence of an 
external magnetic background, we have performed a deeper investigation
of the phenomenon, which has extended the previous analysis 
of Ref.~\cite{Bonati:2014ksa} in various directions.
We have thus studied the static potential 
for various orientations of the quark-antiquark separation with
respect to the background field,
in order to obtain information about its angular dependence.
Then, exploiting new numerical simulations
performed for finer values of the lattice spacing, we have extracted a continuum
extrapolation of our results and tried to extend the analysis towards
larger magnetic fields. As a byproduct of our study, 
we have also obtained a continuum extrapolation of the static
potential at zero external field.
Finally, we have extended our investigation
at finite temperature, exploring the effects of the magnetic 
background below the pseudo-critical temperature $T_c$.
Our main results can be summarized as follows.

At zero temperature, the full angular dependence of the anisotropic
potential can be described assuming that it has a Cornell form  
along each direction. Moreover, the angular dependence of the 
parameters can be accounted for by the first allowed term 
$c_2$ in a Fourier expansion in $\theta$, which corresponds to a quadrupole
deformation.
On the other hand, the continuum extrapolation of our results shows that
the magnetic field induces a significant anisotropy 
only for the string
tension, while the Coulomb coupling is almost unaffected, at least
for $|e|B \lesssim 1$ GeV$^2$. The latter result is compatible with 
an analysis based on the numerical 
study of gluon field strength correlators~\cite{D'Elia:2015dxa}.

The observed anisotropy of the potential  suggests that the 
string tension in the direction parallel to the magnetic field might disappear 
for magnetic
fields of the order of 
$|e| B \sim 4$ GeV$^2$. While this extrapolation to large
magnetic fields cannot be supported by a reliable continuum limit
in that regime, we have verified that it is consistent with results obtained 
at the finest explored lattice spacing, where we managed to perform
numerical simulations for magnetic fields up $|e| B \sim 3$ GeV$^2$.
Future numerical studies on finer lattices could check 
the hypothesis of a possible longitudinal deconfinement 
at large $B$, which presently is just suggested by our results.

Finally, our finite temperature results, obtained in an interval
ranging approximately from 100 MeV to right below the pseudocritical
temperature, have shown that the main effect of the magnetic field on
the static potential in that range consists of a general suppression
of the string tension and of the confinining properties of the
medium. Moreover, such phenomenon happens even when still no
effect of inverse magnetic catalysis is visible in the chiral
condensate, thus suggesting that the decrease of the pseudocritical
temperature as a function of $B$, usually named as ``inverse magnetic
catalysis'', might be understood in terms of a ``deconfinement
catalysis'', representing the leading physical phenomenon: this idea
is also supported by some recent computations 
in holographic models~\cite{Dudal:2015wfn}.
From a phenomenological point of view, 
the precocious disappearance of the confining properties
of the static potental might have a significant influence 
on the suppression of heavy quark bound 
states produced in the thermal medium, even below $T_c$,
in all situations in which a strong magnetic field might be present,
e.g. in non-central heavy ion collisions and in the thermal medium of 
the early Universe. This is an issue of particular interest, which should
be further investigated in future studies.
\\

\acknowledgments

We acknowledge PRACE for awarding us access to resource FERMI based in Italy at
CINECA, under project Pra09-2400 - SISMAF.  FS received funding from the
European Research Council under the European Community Seventh Framework
Programme (FP7/2007-2013) ERC grant agreement No 279757.  FN acknowledges
financial support from the INFN SUMA project.

\end{document}